\documentclass[oldversion]{aa600}
\usepackage{graphicx}
\usepackage{txfonts}
\usepackage{natbib}
\bibpunct{(}{)}{;}{a}{}{,}

\begin{document}

  \title{30 Glitches in slow pulsars}

  \author{G.\ H.\ Janssen\inst{1} 
    \and B.\ W.\ Stappers\inst{2,1}}
      
  \institute{Astronomical Institute ``Anton Pannekoek'', University of
  Amsterdam, Kruislaan 403, 1098 SJ Amsterdam, The Netherlands;\\  \email{gemma@science.uva.nl}
  \and Stichting ASTRON, Postbus 2, 7990 AA Dwingeloo, The
  Netherlands;  \email{stappers@astron.nl}}

  \abstract{We have analyzed 5.5 years of timing observations of 7
  ``slowly'' rotating radio pulsars, made with the Westerbork
  Synthesis Radio Telescope. We present improved timing solutions and
  30, mostly small, new glitches. Particularly interesting are our
  results on PSR~J1814$-$1744, which is one of the pulsars with
  similar rotation parameters and magnetic field strength to the
  Anomalous X-ray Pulsars (AXPs). Although the high-B radio pulsars
  do not show X-ray emission, and no radio emission is detected
  for AXPs, the roughly similar glitch parameters provide us with
  another tool to compare these classes of neutron stars.
  Furthermore, we were able to detect glitches one to two orders of
  magnitude smaller than before, for example in our well-sampled
  observations of PSR~B0355$+$54.  We double the total number of known
  glitches in PSR~B1737$-$30, and improve statistics on glitch sizes
  for this pulsar individually and pulsars in general.  We detect no
  significant variations in dispersion measure for PSRs B1951$+$32 and
  B2224$+$65, two pulsars located in high-density surroundings.  We
  discuss the effect of small glitches on timing noise, and show it is
  possible to resolve timing-noise looking structures in the residuals
  of PSR~B1951$+$32 by using a set of small glitches.

  }

  \date{Received/Accepted}

\maketitle

\section{Introduction}

Two sorts of irregularities, in the otherwise very stable pulsar rotation
rates exist, which limit the accuracy to which pulse arrival times can
be measured: timing noise and glitches.\\
Timing noise is seen as random fluctuations in the rotation rate of
the pulsar on timescales of days to years. It is largest in young
pulsars and pulsars with large period derivatives \citep{ch80,antt94}.

Glitches on the other hand are characterized by a sudden increase of
the pulsar rotation frequency $(\nu)$, accompanied by a change in
spindown rate $(\dot\nu)$ and sometimes followed by relaxation or
exponential decay to the previous rotation state.  Typical magnitudes
of glitches are from $10^{-10}\nu$ to $10^{-6}\nu$ and steps in
slowdown rate are on the order of $10^{-3}\dot\nu$.  Glitches give an
unique opportunity to study the internal structure of neutron stars,
as they are believed to be caused by sudden and irregular transfer of
angular momentum from the superfluid inner parts of the star to the
more slowly rotating crust \citep{rzc98}. They are mostly seen in
pulsars with characteristic ages ($\tau_c$) around $10^4-10^5$ yr and
can occur up to yearly in some pulsars.  Most of the youngest pulsars,
with $\tau_c\lesssim2000$~yr show very little glitch activity. This
could be because they are still too hot which allows the transfer of
angular momentum to happen more smoothly \citep{ml90}.

A lot of new glitches have been found in the last decades
\citep{lsg00,uo99,wmp+00}. This makes statistical analysis of glitch
sizes and activity possible. However, no clear relations between
glitch parameters or dependence of glitch parameters on rotation
properties have been found so far. Extending the sample of glitches,
especially with previously unknown small-magnitude glitches will
lead to a better insight into the glitch mechanism and the
structure of the neutron star.

\section{Observations and data analysis}

We have observed our sample of pulsars since 1999 at the Westerbork Synthesis
Radio Telescope (WSRT) with the Pulsar Machine (PuMa; \citealt{vkh+02}).
Depending on brightness, we observed each pulsar at multiple
frequencies centered at 328, 382, 840 or 1380 MHz as shown in Table
\ref{tab:frequencies}. We have approximately monthly observations of 6
to 60 minutes duration.  The sampling time for all observations was
0.4096~ms and the bandwidth used was 8$\times$10~MHz for the observations at
840 and 1380 MHz. The low-frequency observations only used 10~MHz of
bandwidth, and were usually simultaneous at both frequencies.
Each 10 MHz band was split into 64 channels.

The data was dedispersed and folded offline, and then integrated over
frequency and time over the whole duration to get a single profile for
each observation.  The profile was cross-correlated with a standard
profile, obtained from the summation of high signal-to-noise (S/N)
profiles, to calculate a time of arrival (TOA) for each
observation. These were referred to local time using time stamps from
a H-maser at WSRT.  These times are converted to UTC using GPS maser
offset values measured at the observatory, and clock offsets from the
IERS\footnote{http://www.iers.org}. The times of arrival were
converted to the Solar system barycentre using the JPL ephemeris DE200
\citep{sta82}.

The TOAs are analysed with TEMPO (version
11.002\footnote{http://pulsar.princeton.edu/tempo/}), using a model
including position, rotation frequency, first derivative and
dispersion measure (DM) of the pulsar.  In this version of TEMPO it was
only possible to fit for the parameters of 4 different glitches
simultaneously. Because of the large numbers of glitches we found, in
for example PSR~B1737$-$30, it was necessary to be able to fit for
more glitches, so we extended the number of glitch parameters to enable
fitting up to 9 glitches simultaneously.

A glitch can be recognized by the sudden increase of the rotation
frequency, causing the residuals from the timing model to increase
towards earlier arrival (e.g. Figs.~\ref{fig:0355} and
\ref{fig:1814}).  Glitches can be described as combinations of steps
in rotational frequency and its derivative, of which parts can decay
exponentially on various timescales.  Usually only the larger glitches
show measurable exponential decay. Because we mostly found
small-magnitude glitches, we could only fit for the permanent steps in
frequency and frequency first derivative.  The few large glitches we
found in PSR~B1737$-$30 were not sufficiently well sampled to enable
us to fit the exponential decay, or a new glitch happened too soon
after a large one which caused the relaxation part, if present, to be
absorbed in the next glitch parameters.

Solutions before and after the glitch epoch are used to roughly estimate
the steps in frequency and its first derivative.  If the gap
between observations around the glitch epoch is not too large, the
epoch can be found more accurately by requiring a phase-connected
solution over the gap. This is done using another parameter in TEMPO,
GLPH, which represents a fraction of the phase needed as a jump to get
a connected solution. Minimization of this parameter by changing the
epoch results in the best value for the glitch epoch.  However if the
gap is too large, rotations can be missed and the glitch epoch is
estimated as the mid-point of the gap.

\begin{table}[b]%
  \centering
  \caption{Summary of observed frequencies.
  \label{tab:frequencies}}%
  \begin{tabular}{lcccc}%
    \hline
    \hline
    Pulsar Name  &  328 MHz  & 382 MHz  &  840 MHz  &  1380 MHz  \\
    \hline
    B0355$+$54 & $\star$ & $\star$ & $\star$ & $\star$\\
    B0525$+$21 & $\star$ & $\star$ & $\star$ & $\star$\\
    B0740$-$28 & $\star$ & $\star$ & $\star$ & $\star$\\
    B1737$-$30 &  &  & $\star$ & $\star$\\
    J1814$-$1744 &  &  &  & $\star$\\
    B1951$+$32 &  &  & $\star$ & $\star$\\
    B2224$+$65 &  &  & $\star$ & $\star$\\
    \hline
  \end{tabular}
\end{table}

\section{Individual pulsars: Results}

\begin{table*}
  \centering
  \caption{Position and rotation parameters. The uncertainties are the
  standard TEMPO errors and refer to the last digit quoted. Rms values
  result from models including glitch parameters presented in Table \ref{tab:glitch}.
    \label{tab:parameters}}
  \begin{tabular}{lccc}
    \hline
    \hline
    Pulsar name     & B0355$+$54        & B0525$+$21     & B0740$-$28 \\
                    & J0358$+$5413      & J0528$+$2200   & J0742$-$2822\\
    \hline
    Right ascension~(J2000)      & 03:58:53.7166(3)  & 05:28:52.26(1) & 07:42:49.082(3)\\
    Declination~(J2000)     & 54:13:13.717(3)   & 22:00:03(3)    & $-$28:22:43.96(6)\\
    Epoch~(MJD)     & 52350             & 52362          & 52200   \\
    Data span~(MJD) & 51386$-$53546     & 51408$-$53548  & 51386$-$53546\\
    Frequency $\nu$~(s$^{-1}$)  & 6.39453829911(4)   & 0.26698470619(3)& 5.996409260(2)\\          
    Frequency first derivative $\dot\nu$~($10^{-15}$\,s$^{-2}$)
                    & $-$179.747(1)       & $-$2.8537(3)     &  $-$604.71(4)   \\
    Frequency second derivative $\ddot\nu$~($10^{-24}$\,s$^{-3}$)
                    & 0.18(2)           &                & 0.9(1) \\
    Dispersion measure $DM$~(pc\,cm$^{-3}$)& 57.1420(3)        & 50.87(1)       & 73.782(2)\\
    Characteristic age $\tau_c$~(kyr)     & 564               & 1480           & 157  \\
    Surface magnetic field strength $B$~($10^{12}$G)  & 0.84              & 12.4           & 1.69 \\
    Glitch activity $A_g$~($10^{-7}$\,yr$^{-1}$) & 0.0004    & 0.003          & 0.01 \\
    Number of TOAs  & 184               & 83             & 209 \\
    Rms timing residual~($\mu$s)   & 47                & 777            & 709 \\
  \hline
  \end{tabular}
\end{table*}

\begin{table*}%
  \centering
  \begin{tabular}{lcccc}%
    \hline
    \hline
              & B1737$-$30  &               & B1951$+$32 & B2224$+$65\\
              & J1740$-$3015& J1814$-$1744  & J1952$+$3252 & J2225$+$6535\\
    \hline
    Right ascension~(J2000) &17:40:33.75 & 18:14:43.10(5) & 19:52:58.164(5)& 22:25:52.627(3)\\ 
    Declination~(J2000)&$-$30:15:43.8 &$-$17:44:48(8)& 32:52:40.38(5) & 65:35:35.16(2)\\
    Epoch~(MJD)&52200       &52353           & 52245         & 52400\\
    Data span~(MJD)& 51427$-$53100 & 51391$-$53545 & 51427$-$53546 & 51427$-$53546\\
    Frequency $\nu$~(s$^{-1}$) &1.6480330270(4)& 0.251515052(1)&25.29570396(1)&1.4651138511(2)\\
    Frequency first derivative $\dot\nu$~($10^{-15}$\,s$^{-2}$)
                   &$-$1266.635(7)& $-$47.11(2) & $-$3737.3(4)& $-$20.774(3)\\
    Frequency second derivative $\ddot\nu$~($10^{-24}$\,s$^{-3}$) &  &  & 54(8)      & $-$0.55(3)\\
    Dispersion measure $DM$~(pc\,cm$^{-3}$)& 152.15(2)& 830           & 45.19(2)      & 36.42(2)\\
    Proper motion in right ascension~(mas\,yr$^{-1}$)&  &         &              & 106(15) \\
    Proper motion in declination~(mas\,yr$^{-1}$)&  &     &              & 174(18) \\
    Characteristic age $\tau_c$~(kyr)      & 21   & 85              & 107          & 1118\\
    Surface magnetic field strength $B$~(10$^{12}$G)   & 17   & 55      & 0.49         & 2.6\\
    Glitch activity $A_g$~($10^{-7}$\,yr$^{-1}$) & 1.6 & 0.08      & 0.008        & 0.0007 \\
    Number of TOAs    & 64         & 80              & 106          & 97   \\
    Rms timing residual~($\mu$s)&150       & 8580            & 359         & 215 \\
  \hline 
  \end{tabular}
\end{table*}

\subsection{PSR~B0355$+$54 (J0358+5413)}

This relatively old pulsar ($5.6\times 10^5$ yr) has been studied
intensively since its discovery \citep{mth72}.
It has low timing noise and the only report of
glitches in this pulsar is by \cite{lyn87} and \citet{sha90}.  The
reported glitches are very different, the first happened at MJD 46079
and was quite small with $\Delta\nu/\nu=5.6\times 10^{-9}$. The
second, at MJD 46496, is one of the largest glitches known, with
$\Delta\nu/\nu=4.4\times 10^{-6}$.

\begin{figure}
  \centering
  \includegraphics[width=7.5cm]{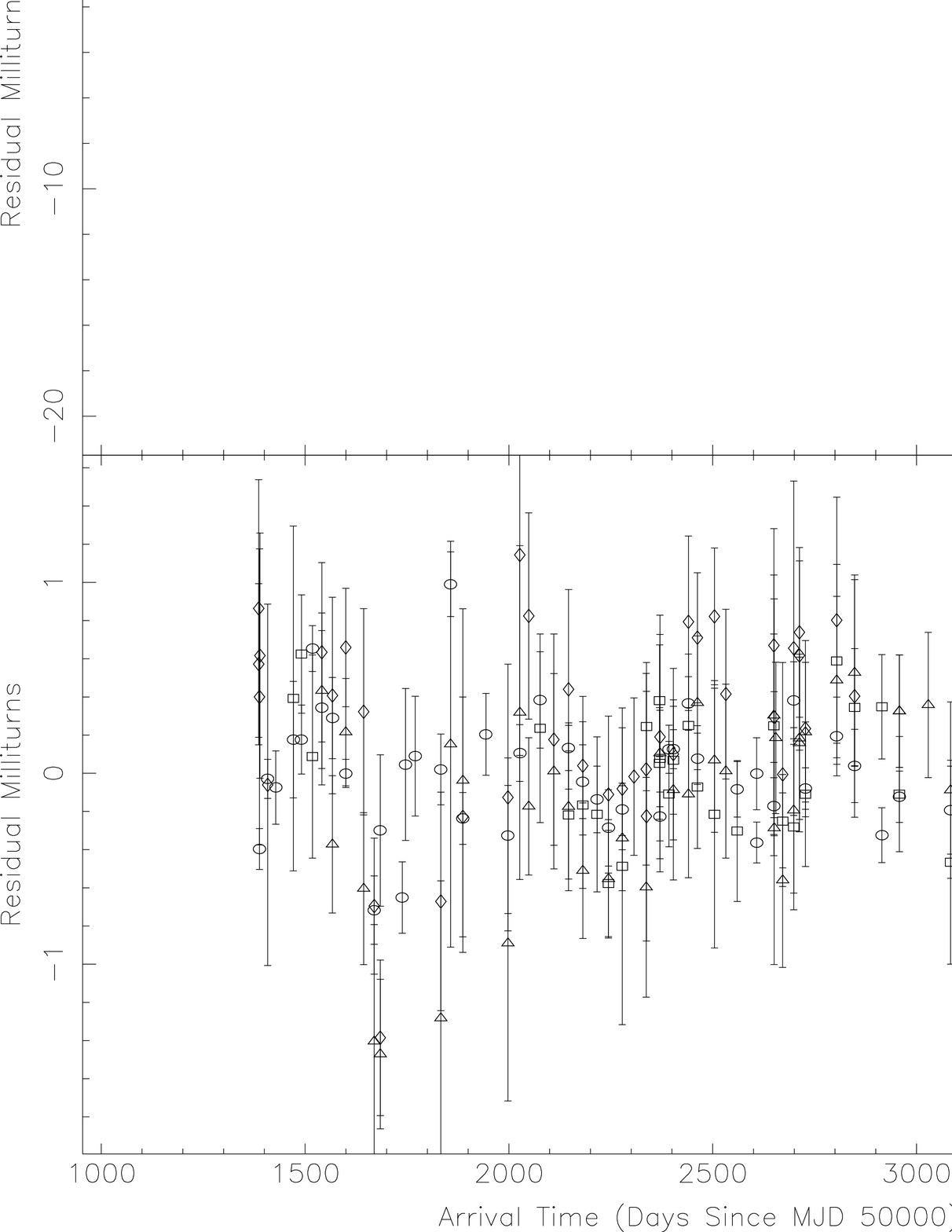}
  \caption{Detection of four miniglitches in PSR~B0355$+$54.  a)
   Residuals when not fitting for any glitches and keeping
   best-solution rotation and position parameters fixed
   (Table \ref{tab:parameters}). The epochs of the glitches are
   indicated by arrows in the upper plot. The residuals bend down to
   earlier arrival at the epoch of the first glitch (MJD~51673). b)
   Like plot (a), but also including the parameters for the first
   glitch; the residuals bend down at the epoch of the second glitch
   (MJD~51965). c) Third glitch (MJD~52941). d) Fourth glitch
   (MJD~53216). Plot (e) shows the residuals for our best-fit model,
   including the four mini-glitches (Table~\ref{tab:glitch}).
  \label{fig:0355}}
\end{figure}

We have a well-sampled multi-frequency observing data span of almost 6
years for this pulsar. We improve on previously published
\citep{hlk+04,wmz+01} timing solutions. Although a timing solution for
our data span including only position parameters and two frequency
derivatives already yields a better root-mean-square (rms) of 67~$\mu
s$, we find a better solution (47~$\mu s$) including 4 mini-glitches
with frequency steps over an order of magnitude smaller than any
glitch found to date in a slowly rotating pulsar: $\Delta\nu/\nu$ of
$10^{-10}$ to $10^{-11}$.  In Fig.  \ref{fig:0355} the residuals with
and without fitting for these glitches are shown. The upper plots show
the result for keeping all parameters fixed but leaving the glitches
out one at the time, the bottom plot shows the best-fit solution to
our data including the four mini-glitches. Parameters for this
solution can be found in Tables \ref{tab:parameters} and
\ref{tab:glitch}.

Because these mini-glitches are so much smaller than any glitch found
before, we investigated what our detection limit is for glitches in
this pulsar by carrying out simulations.  We first created a set of
times of arrival resembling a model including a mini-glitch. This was
done by fitting our data set to the model, but keeping all, including
the simulated glitch parameters, fixed. The resulting residuals were
used to correct all individual TOAs so that they would follow the
model exactly.  Then random noise was added to each TOA with a level
typical for this pulsar.  These new TOAs were used in the usual
process of finding a timing solution to see whether we could
significantly redetect the glitch.  For this pulsar, the simulations
have shown that we can detect glitches as small as $\Delta\nu/\nu=
10^{-11}$.

\subsection{PSR~B0525$+$21 (J0528+2200)}

Only one small glitch has been observed before in this old, very
slowly rotating pulsar \citep{dow82, sl96}.
 It occurred at MJD 42057 and had a magnitude of
$\Delta\nu/\nu=1.2\times~10^{-9}$.  An exponential decay of 50\% in
150 days was measured.

We find a glitch at MJD 52284 with similar parameters as the one
before, with a magnitude of $\Delta\nu/\nu=1.46\times~10^{-9}$.
Unfortunately the new glitch occurs in a large gap in our data span
which makes it impossible to fit for an exponentional decay.  Our data
suggests that another glitch happened around MJD 53375 (January
2005). When using data up until June 2005, we can only fit for the
step in frequency, which appears to be small, $\Delta\nu/\nu=1.7\times~
10^{-10}$.

\subsection{PSR~B0740$-$28 (J0740-2822)}

\cite{dmk+93} have reported two small glitches in this pulsar, at
MJDs 47678 and 48349 with relative sizes 2.63 and 1.5
$\times~10^{-9}$. Another two small glitches (1.9 and
1.3~$\times~10^{-9}$) at MJDs 49298 and 50376 were reported by
\cite{uo99}.

With the four new glitches we find, this pulsar is showing regular
glitching behaviour, with small glitches roughly every 900 days. All
glitches are surprisingly similar in magnitude.

\subsection{PSR~B1737$-$30 (J1740-3015)}

This frequently glitching pulsar has been intensively studied since
1987 by \cite{ml90}, \cite{sl96}, \cite{uo99} and \cite{klg+03}.  It has a large
spindown value of $-1.267\times~10^{-12}$ s$^{-2}$.  Large glitches,
with magnitudes $\Delta\nu/\nu$$\approx$$10^{-7}$, occur roughly every
800 days, and smaller ones every 2--300 days.  Fourteen glitches have been
reported up to MJD 51000.

Because of the large number of glitches in a short time, and the
location of the pulsar close to the ecliptic plane, it is difficult to
find a good value for the pulsar declination in the timing
solution. Therefore we first use a recently published value for the
position, from \cite{wmz+01}, and keep this position fixed when
fitting for the glitches.

Due to a limited number of data points in the middle of our data set,
we were at first not able to find a connected timing
solution.  In overlap with our data span, five glitches were reported
in the conference proceedings of the Lake Hanas Pulsar
meeting\footnote{http://www.uao.ac.cn/hanas/presentation.htm} (2005)
by Esamdin et al. and Zou et al.  We used the glitch parameters they
found to create a phase-connected solution for our data set covering
almost 4.5 years.

We find a solution confirming those glitches and add 5 new small
ones, which results in a continuing high glitch rate of $10$ new
glitches in 4.5 years for this pulsar. Some of those are an order of
magnitude smaller than the smallest glitch found before for this
pulsar.

We almost double the total number of glitches known for this
pulsar. This allows us to have better statistics on the glitch size
distribution.  In the bottom panel of Fig. \ref{fig:glitches2} a
histogram of glitch sizes is shown, including all known glitches for
PSR~B1737$-$30.  With the new, mostly small glitches included, the
sizes appear to be roughly equally distributed between
$\Delta\nu/\nu=10^{-10}$ and $10^{-6}$.

\subsection{PSR~J1814$-$1744}
This pulsar was discovered in 1997 \citep{ckl+00}. Its estimated
surface dipole magnetic field strength is one of the highest known for
radio pulsars, $5.5\times~10^{13}$G. This is very close to the magnetic
field strengths for anomalous X-ray pulsars (AXPs). The spin
parameters are very similar as well.  However, no X-ray emission was
detected for this pulsar \citep{pkc00}.  There are now six high-B
radio pulsars known.  Two have been detected in X-rays, but all have
X-ray luminosity upper limits one to three orders of magnitudes less
than AXPs \citep{pkc00,msk+03,gklp04,km05}.

Three glitches have been detected in AXPs so far: two in
1RXS~J1708$-$4009, and one in 1E~2259$+$586. In Table \ref{tab:axp} the
glitches found for all high-magnetic field objects are shown.  The
steps in frequency in AXP glitches seem at least an order of magnitude
larger than those in the high-B radio pulsars, but the values are not
uncommon for radio pulsars in general. For example the Vela pulsar and
PSR~B1737$-$30 show comparably large-magnitude glitches.

\begin{figure}
  \centering
    \includegraphics[width=8cm]{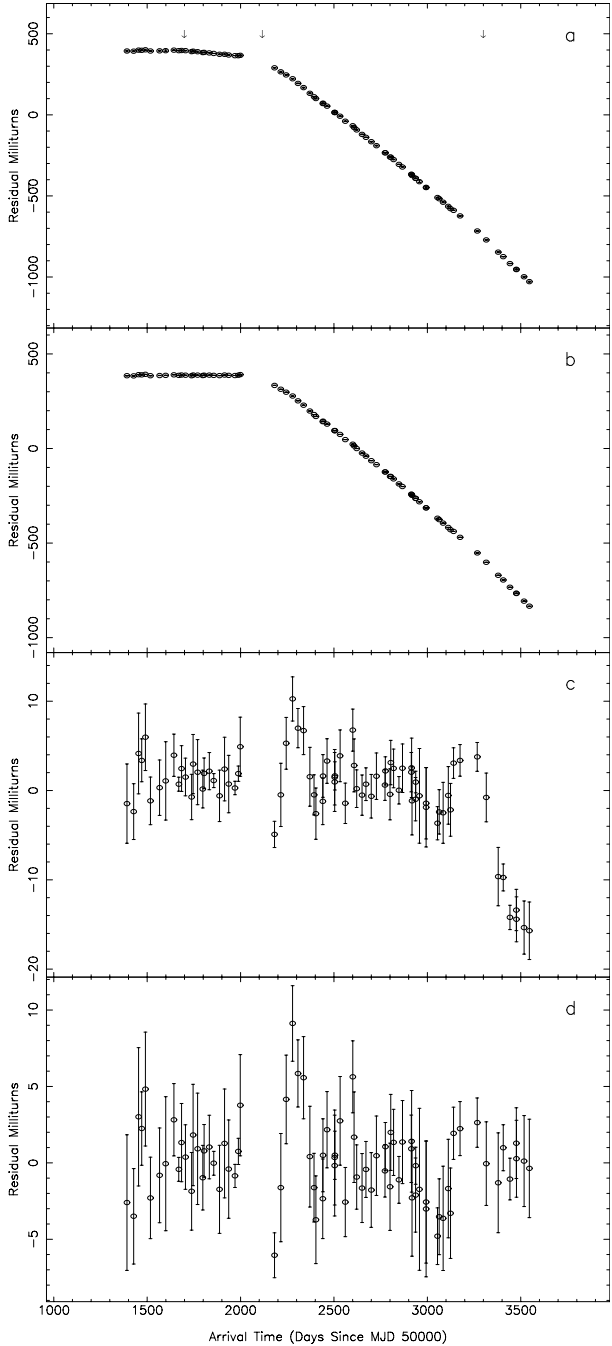}  
  \caption{Glitch detections in PSR~J1814$-$1744. Like
  Fig.~\ref{fig:0355}, the upper plots (a,b,c) show the residuals for the
  best-fit model and including parameters for each glitch in turn for
  each following plot. The second, larger glitch is already clearly
  visible in plot (a). Plot (d) shows the residuals for the best-fit model
  including the parameters for the three glitches detected.
  \label{fig:1814}}
\end{figure}

We have detected three glitches for PSR~J1814$-$1744. The parameters are
listed in Table \ref{tab:glitch} and the residuals are shown in Fig.
\ref{fig:1814}.  Together with another glitch detected in the
high-magnetic field pulsar J1119$-$6127 \citep{ckl+00}, we are
now able to compare AXPs and radio pulsars in another way.

\subsection{PSR~B1951$+$32 (J1952+3252)}

This low-magnetic field pulsar was discovered in 1987 by
\cite{kcb+88}, and is associated with the CTB 80 supernova remnant
\citep{str87}.  Timing solutions have been presented by \cite{flsb94}
and \cite{hlk+04}. A small glitch was detected by \cite{fbw90} in the
beginning of March 1988.

Because of the high timing activity of this pulsar, it was quite
difficult to find a new timing solution for this pulsar when starting
from a solution with an epoch just before the start of our data
span. Only by shifting the epoch a few hundred days at a time and
creating new solutions for each next epoch could we generate the
present solution with the epoch in the middle of our data set.

\begin{figure}
  \centering
 \includegraphics[width=8cm]{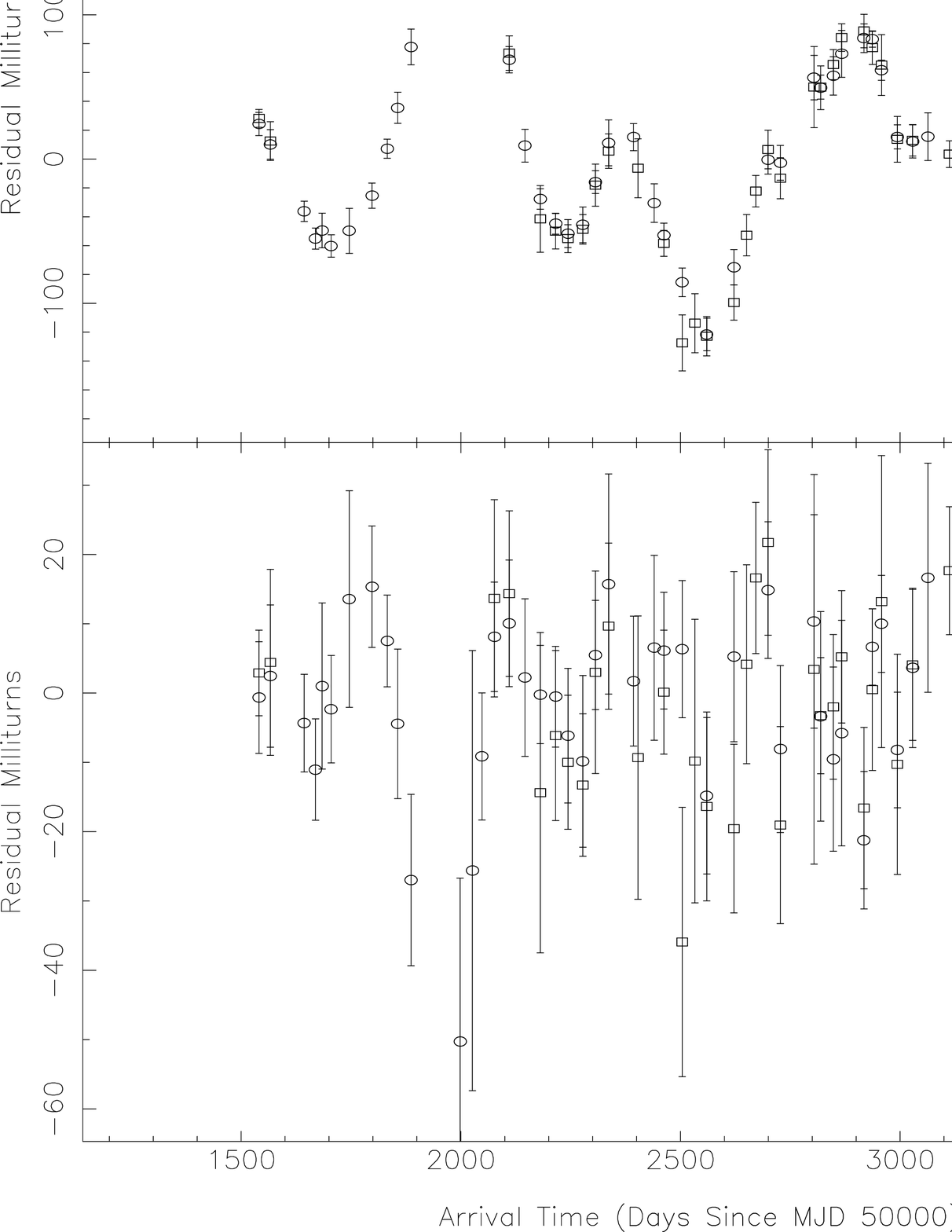}  
  \caption{PSR~B1951+32: a) Residuals to a model including two
  frequency derivatives. b) Residuals for a timing
  solution also including a third frequency derivative. c) Residuals
  for the solution presented in Tables~\ref{tab:parameters} and
  \ref{tab:glitch}. Note that the vertical scale is $10$ times smaller
  than plots (a) and (b). For all diagrams, circles are observations at
  1380 MHz, squares at 840 MHz.
  \label{fig:1951}}
\end{figure}

We found a solution for our data span of 5.5 years consisting of
observations at 840 and 1380 MHz. The solution has a rms of 3.2 ms,
which is the best timing solution so far found for this pulsar only
including the first two frequency derivatives.  The residuals show a
large timing activity, see Fig. \ref{fig:1951}-a.  We noticed that
the $\ddot\nu$ was surprisingly stable when fitting for small parts of
the data, therefore we also included a third derivative in our
solution.  This results in a better solution (2.2 ms) than the one
using only two frequency derivatives, but still a lot of timing
activity is present (Fig. \ref{fig:1951}-b).  

\begin{figure}
  \centering
  \includegraphics[width=6cm,angle=270]{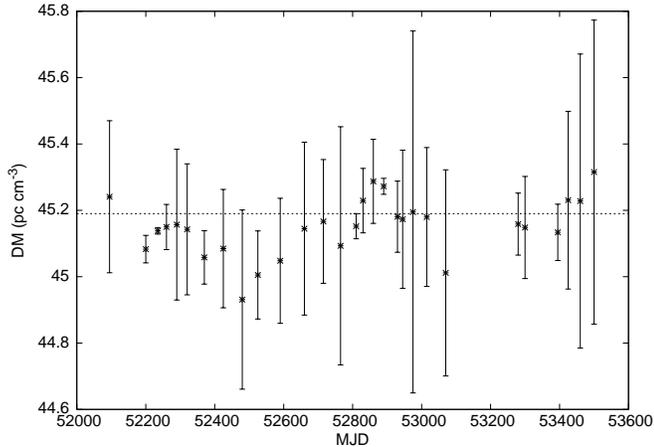}
  \caption{DM measurements for PSR~B1951+32. Each point is calculated
  using two observations at 1380 MHz, and two observations at 840
  MHz. For each point, we fit for DM keeping the parameters for the
  solution presented in Fig.~\ref{fig:1951}-a
  fixed. The horizontal line represents the average DM found for the
  best timing solution.
  \label{fig:dmvar}}
\end{figure}

The pulsar is currently interacting with its supernova remnant, and
has a high proper motion \citep{hllk05}.  Therefore it is not unlikely
that DM variations are influencing our timing solution.  To get a
better view of the influence of the medium surrounding the pulsar on
the arrival times, we searched for dispersion measure variations in
our data set. Starting from the solution including only position and
rotation parameters, we fixed all parameters and fitted only for DM
over small parts of the data set. The result can be seen in Fig.
\ref{fig:dmvar}.  Although our errors are too large to constrain the
variations, the DM appears to be wandering slowly around the mean
value of $45.19$ pc cm$^{-3}$, but no clear trend is seen. This
indicates that DM variations are not responsible for the timing-noise
in this pulsar.

The pulsar has shown glitching behaviour before, and the cusp-like
structures in the residuals are known to be an indication for glitches
\citep{hob02}. Therefore we tried to resolve the variations with
glitches.

A solution including four glitches results in a much better rms of 0.4
ms for our data span. The glitches we use are of similar magnitude to
the one reported by \cite{fbw90}. The steps in frequency derivative
are also comparable, see Table~\ref{tab:glitch}. The residuals are
shown in Fig. \ref{fig:1951}-c. Another glitch is likely to have
happened around MJD 51500, at the beginning of our observing
span. Because we have only three data points before that date, which
prevents us from sampling the rotation properties properly, we
have excluded them.

\subsection{PSR~B2224$+$65 (J2225+6535)}

For this old pulsar only one glitch has been found before, by
\cite{btd82}. It was very large, with a magnitude of
$\Delta\nu/\nu\approx1.7\times10^{-6}$.  An upper limit of
$\Delta\dot\nu/\dot\nu=0.003$ on the step in frequency derivative
was given.  \cite{sl96} used improved rotation parameters to confirm
the glitch parameters, and they also report a $\ddot\nu$ of
$0.1\times10^{-24}$s$^{-3}$.

We report three very small new glitches for this pulsar, with steps in
rotation frequency of $\Delta\nu/\nu$ of $0.14, 0.08$, and
$0.19~\times~10^{-9}$ respectively (Table \ref{tab:glitch}).

PSR~B2224$+$65 is traveling at very high speed in the Guitar
Nebula. \cite{cc04} report on shape evolution of the tip of the
nebula, where the pulsar is located: ''the observations reflect the
motion of the pulsar through random density inhomogeneities combined
with a gradient toward a region of lower density.''

\begin{figure}
  \centering
  \includegraphics[width=6cm,angle=270]{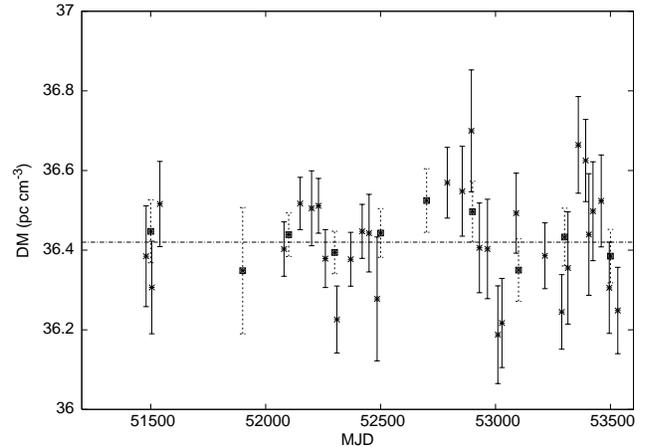}
  \caption{DM variations for PSR~B2224+65. Crosses: each
 point is calculated using two observations at 1380 MHz, and two
 observations at 840 MHz. Squares: each point is calculated using all
 data points in intervals of 200 days. Horizontal line is the average
 DM value for the best timing solution.
  \label{fig:2224}}
\end{figure}

Like for PSR~B1951$+$32, we have searched our data for dispersion
measure variations. The result is shown in Fig.~\ref{fig:2224}. Our
data is consistent with a stable dispersion measure. However, we
have insufficient sensitivity to measure the very small variations
predicted by \cite{cc04}.

\section{Discussion}

\subsection{Glitch sizes}

Little is known about the range of glitch sizes and their distribution
within this range.  \cite{lsg00} showed a distribution for 48 glitches
in 18 pulsars. They claimed there was a peak around
$\Delta\nu/\nu\approx10^{-6.5}$, but they also noted that the glitch
sample could be incomplete at the lowest level.  How small do we
expect glitches to be?  Glitches are believed to be the result of the
unpinning of neutron superfluid vortices which causes transfer of
angular momentum of the interior superfluid to the solid crust of the
star.  There is no report of restrictions on a minimum stress needed
before vortices unpin, so it is not unlikely that the observed lower
limit on glitch sizes is mainly due to the detection limits of the
observing systems and the timing precision achievable for a given
pulsar.

In this paper we measure glitch sizes down to
$\Delta\nu/\nu=10^{-11}$, which provides the first evidence that such
small glitches occur and can be measured in slowly rotating pulsars.
These glitches are then of a similar size to the one reported by
\cite{cb04} in a millisecond pulsar, and thus perhaps provide further
evidence for a continuous distribution of glitch sizes.

\begin{figure}
  \centering
  \includegraphics[width=6cm,angle=270]{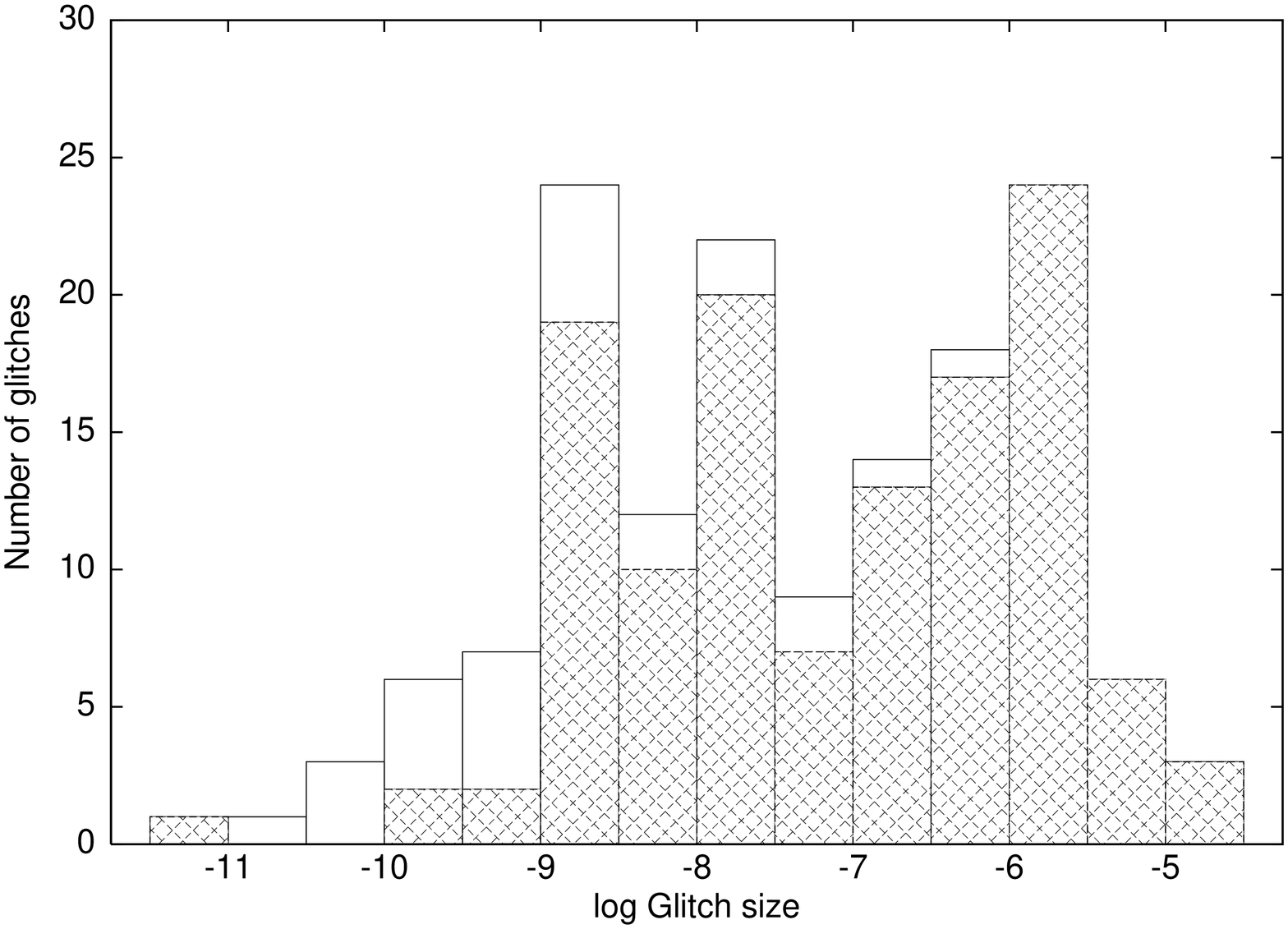}  
 
  \includegraphics[width=6cm,angle=270]{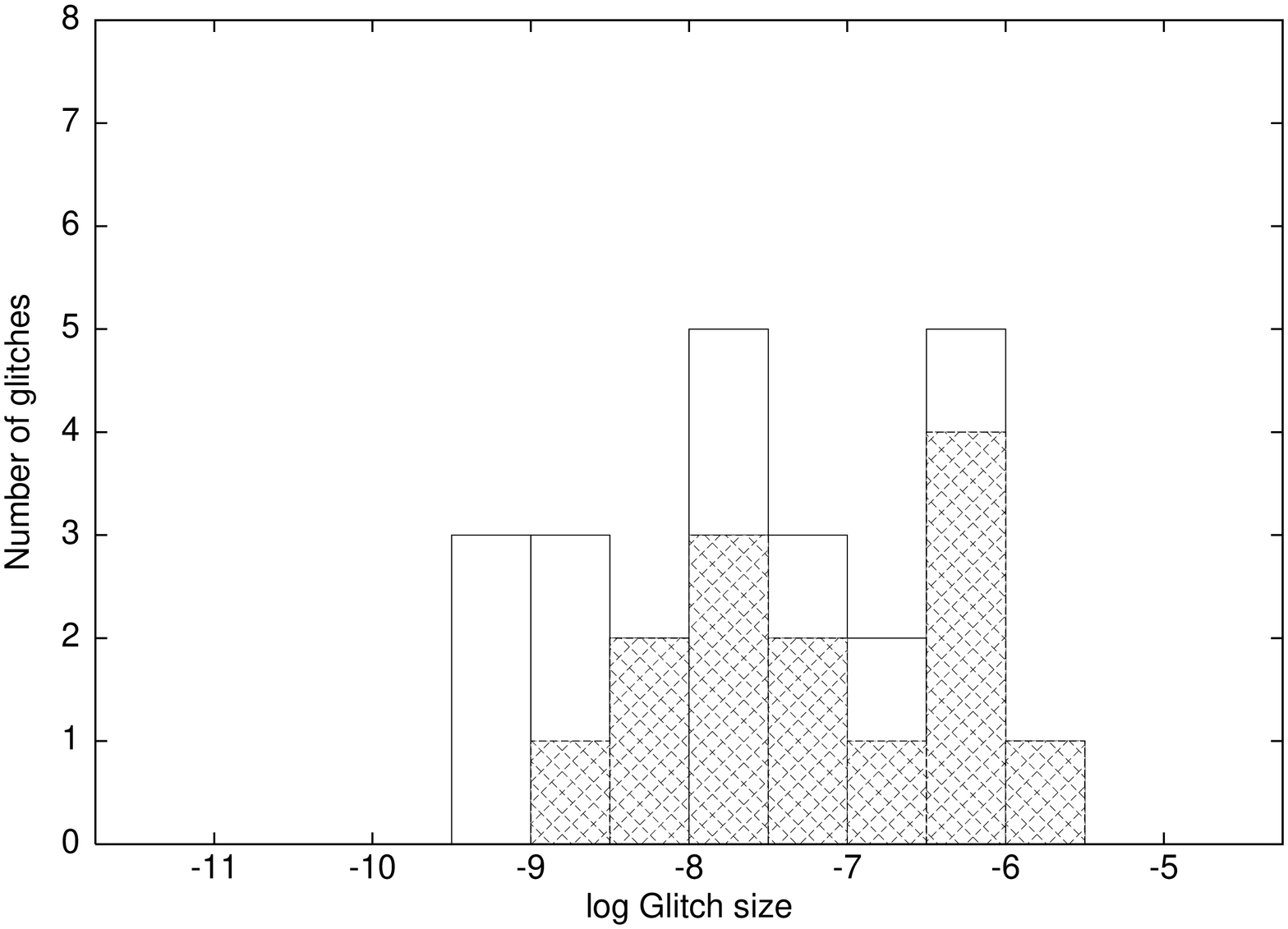} 

  \caption[]{Upper panel: histogram of all known glitch sizes, from
  the ATNF glitch table\addtocounter{footnote}{-1}\footnotemark\, \citep{mhth05} and
  \cite{uo99}. New glitches found in this study are added on top of
  the known glitches.  The lower panel shows the glitches of
  PSR~B1737$-$30.  
  \label{fig:glitches2}}
\end{figure}

Let us now consider how these small glitches affect the observed
glitch size distribution.  In Fig. \ref{fig:glitches2}, a histogram
is shown for all now known glitch sizes.  New glitches found in this
study are shown added on top of the old glitch distribution.  
A Kolmogorov-Smirnov test shows that over the whole range, the
distribution has only a probability of $0.001$\% to be
consistent with a flat distribution
  in log space of glitch sizes. But if we
consider only the part of the diagram between
$10^{-9}~<~\Delta\nu/\nu~<~10^{-5.5}$, where the statistics are better,
another KS test shows that the distribution has a $29.8$\% chance
to be drawn from a flat distribution.  The increased number of
glitches with sizes around $\Delta\nu/\nu\approx10^{-9}$, now
comparable to the amount of larger glitches observed, suggests again
that the lack of the smallest glitches at the lower end of the
distribution is due to observing limits.  The lack of glitches at the
upper end of the distribution can not be due to observing
limits. Apparently there is some physical restriction to the maximum
size of a glitch, and we can consider the boundary of
$\Delta\nu/\nu\approx10^{-5}$ as the natural upper limit of glitch
sizes.

The glitch size distribution may be biased because there is a whole
range of different pulsars included,  which can all show different
glitching behaviour. For example the Vela pulsar suffers from a large
number of huge glitches. The fact that those are more easily
detectable, will have a large influence on the overall glitch
size distribution.  We take a frequently glitching pulsar,
PSR~B1737$-$30 and make a similar histogram as before to see what
happens if we study the glitch size distribution for one pulsar. The
result is shown in the bottom panel of Fig. \ref{fig:glitches2}.
As we now have almost doubled the number of glitches detected for this
pulsar, and we measure glitches an order of magnitude smaller than
before, we can do a statistical analysis of the glitch
size distribution for a single pulsar with glitches covering almost
the whole range of sizes observed. Using a Kolmogorov-Smirnov test,
we calculated that in the observed size range there is a $90.2$\%
probability that the glitches of PSR~B1737$-$30 are drawn from a flat
distribution (in log space) of glitch sizes.

The greater glitch rate of PSR~B1737$-$30 allows us to obtain
sufficient glitches to show that the distribution of glitches in this
pulsar are likely drawn from a flat log-space distribution of glitch
sizes. Whether this is representative of the glitch size distribution
of pulsars in general is more difficult to determine based on just
this pulsar and a larger sample of pulsars where such fits can be
determined is required.

\addtocounter{footnote}{1}

\footnotetext{http://www.atnf.csiro.au/research/pulsar/psrcat/glitchTbl.html}

\subsection{Glitch activity}

An indication of glitch activity was first introduced by \cite{ml90}
as the mean fractional change in the rotation frequency per year due
to glitches:
\begin{equation}
A_g=\frac{1}{t_g}\sum\frac{\Delta\nu}{\nu},
\end{equation}
where $\Sigma \frac{\Delta\nu}{\nu}$ is the sum of all glitches
occurring in the observed time span $t_g$, which is measured in
years. They found that the glitch activity is highest for pulsars
with low characteric ages and high frequency derivatives. Most young
pulsars ($\tau_c \lesssim 2000$ yr) show no glitch activity, which is
believed to be an effect of higher temperatures reducing the effect of
vortex pinning.  Apparently for older pulsars, the glitch activity is
higher for pulsars with higher spindown rates (e.g. \cite{wmp+00} and
\cite{uo99}).

Our values for glitch activity for pulsars B0740$-$28 and B1737$-$30
are the same as given by \cite{uo99}, see Table~\ref{tab:parameters}.
However, due to the four mini-glitches in PSR~B0355$+$54 and no large
glitches in our data span, we calculate a much lower value for the
glitch activity of this pulsar than the previously published value
including the large glitch seen in this pulsar.  Although we find
quite a large number of small glitches in the older pulsars of our
sample, the derived glitch activity is still low.

\subsection{Glitch vs timing noise}

Apart from glitches, irregularities in the rotation of the pulsar are
usually described as timing noise. Like glitches, timing noise is also
seen mostly in the younger pulsars with high spin frequency
derivatives.  Timing noise is seen as random wanderings in the
rotation rate and the timescales range from months to years.
Relaxations after glitches are also supposed to be on those
timescales. As we are now able to measure smaller and smaller
glitches, although not yet including relaxations, it is important to
investigate whether small glitches can mimic timing noise.
 
\cite{antt94} use the parameter $\Delta_8$ to quantify the amount of
timing noise in a pulsar. They define the parameter as
\begin{equation}
\Delta(t) = \log \left(  \frac{1}{6\nu}|\ddot\nu|  t^3 \right), 
\end{equation}
which is usually quoted as $\Delta_8$ using a standard time interval
of $10^8$ seconds. They note that values of $\Delta_8$ are expected to
vary for non-overlapping data sets. Therefore it is difficult to make
a statement about the influence of small glitches on this parameter.
What we can do, is compare the value of this parameter for our own
data sets, using solutions with glitches and without glitches. If the
parameter stays the same, we can conclude that it is not influenced by
glitches and the glitches are a different phenomenon than timing noise.
Unfortunately, the variation in $\Delta_8$ is so large, even for
adjacent or partially overlapping intervals of $10^8$ seconds in our
data, that it is impossible to draw a conclusion from the values.

To make a better distinction, if possible, between timing noise and
glitches, more modelling is needed, both on the expected glitch size
distributions, as well as on the exact influence on timing parameters
of small glitches and recoveries from large glitches.  We have seen
that for frequently glitching pulsars, it can be difficult to resolve
glitches that occur close together in time.  This effect is probably
more important for small glitches, as they appear to occur more often
and thus are more likely to merge together.  
There are many manifestations of timing noise and some have a form
which clearly cannot be explained as being due to glitches. However our
discovery of small glitches and the way in which we were able to
improve a ''timing-noise-like'' set of residuals for PSR B1951$+$32 by
including glitches in the solution indicates that they may play a
role, and that improved sensitivity and more frequent observations may
be required to find more such instances.

\begin{table}%
  \caption{Glitch parameters for the 30 new glitches found in this
  study. Where no error in the glitch epoch is quoted, it was not
  possible to constrain the epoch by using the extra glitch parameter
  GLPH. In these cases the middle of the interval between adjacent
  datapoints is quoted as the glitch epoch.
    \label{tab:glitch}}%
  \begin{tabular}{llll}%
    \hline
    \hline Pulsar name & Epoch (MJD)& $\Delta\nu/\nu (10^{-9})$ &
    $\Delta\dot\nu/\dot\nu (10^{-3})$ \\ \hline 
    B0355$+$54 &  51673(15) & \phantom{00}0.04(2) & \\
               & 51965(14) & \phantom{00}0.030(2)& \phantom{00}$-$0.102(7) \\
               & 52941(9) & \phantom{00}0.04(1) & \phantom{$-$00}0.13(4) \\
               & 53216(11) & \phantom{00}0.10(2) & \phantom{00}$-$0.03(4) \\

    B0525$+$21  & 52289(11)     & \phantom{00}1.46(5) & \phantom{$-$00}0.6(1) \\
                & 53379         & \phantom{00}0.17(5) & \\
    
    B0740$-$28  & 51770(20)     & \phantom{00}1.0(3) & \phantom{$-$00}0.9(2) \\
                & 52027(5)      & \phantom{00}2.1(2) & \phantom{00}$-$1.1(2) \\
                & 53090.2(2.6)  & \phantom{00}2.9(1) & \phantom{$-$00}0.39(3) \\
                & 53469.7(8.1)  & \phantom{00}1.1(2) & \\
    
    B1737$-$30  & 51685(21)     & \phantom{00}0.7(4) & \phantom{$-$00}0.09(7)\\
                & 51822(7)      & \phantom{00}0.8(3) & \phantom{00}$-$0.07(7)\\
                & 52007(6)      & \phantom{00}0.7(1) & \phantom{00}$-$0.08(2)\\
                & 52235         & \phantom{0}42.1(9)  &  \\
                & 52271         &           444(5)   & \\
                & 52344         &          220.6(9)  & \\
                & 52603(5)      & \phantom{00}1.5(1) & \phantom{00}$-$0.60(3)\\
                & 52759(5)      & \phantom{00}1.6(3) & \phantom{00}$-$0.59(8)\\
                & 52859         & \phantom{0}17.6(3) & \phantom{$-$00}0.9(1)\\
                & 52943.5       & \phantom{0}22.1(4) &       \\

    J1814$-$1744 & 51700(16)    & \phantom{00}5(2) &  \\
                 & 52117(6)     & \phantom{0}33(2)& \phantom{00}$-$0.5(4)\\
                 & 53302(22)    & \phantom{00}7(2) & \phantom{00$-$}2(1)  \\

    B1951$+$32  & 51967(9)      & \phantom{00}2.25(9) & \phantom{00}$-$0.2(1) \\
                & 52385(11)     & \phantom{00}0.72(9) & \phantom{00}$-$0.04(8)\\
                & 52912(5)      & \phantom{00}1.29(7) & \phantom{00$-$}0.30(9) \\
                & 53305(6)      & \phantom{00}0.51(9) & \phantom{00$-$}0.11(7) \\

    B2224$+$65  & 51900    & \phantom{00}0.14(3) & \phantom{00}$-$2.9(2) \\
                & 52950& \phantom{00}0.08(4) & \phantom{00}$-$1.4(2) \\
                & 53434(13)& \phantom{00}0.19(6) &   \\

    \hline
  \end{tabular}
\end{table}

\begin{table}
  \caption{Glitches in AXPs and high-B radio pulsars. The differences
    in  $\Delta\dot\nu/\dot\nu$ for the second glitch in
    1RXS~J1708$-$4009 are due to a different time scale fitted in the
    papers. \cite{dis+03}
    fit short term change, \cite{kg03} long-term. See Fig.1b of
    \cite{kg03}.\newline References: 1. \cite{dis+03}, 2. \cite{kg03}
    3.\cite{kgw+03}, 4. \cite{ckl+00}, 5. this paper.
   \label{tab:axp}}
  \begin{tabular}{lllll}
    \hline
    \hline
    AXP name & Epoch (MJD) &$\Delta\nu/\nu~(10^{-9})$ & $\Delta\dot\nu/\dot\nu~(10^{-3})$& ref.\\
    \hline
    1RXS          & 51459.0 & \phantom{0}643(5) &\phantom{$-$00}17.2(5) &1\\
     J1708$-$4009 & 51444.601 & \phantom{0}549(20) & \phantom{$-$00}10.0(4) & 2\\
                  & 52015.65& \phantom{0}330(20) &\phantom{$-$0}330(30)$^a$ &1\\
                  & 52014.177 & \phantom{0}141(27) & \phantom{$-$000}0.1(4)$^a$ & 2\\
    \hline
    1E 2259$+$586 &52443.9& 4100(30) & \phantom{$-$}1110(70) & 3\\
    \hline\hline
    Pulsar name & Epoch (MJD) &$\Delta\nu/\nu~(10^{-9})$ & $\Delta\dot\nu/\dot\nu~(10^{-3})$& ref.\\
    \hline
    J1119$-$6127 & 51398 &\phantom{000}4.4(4) &\phantom{$-$000}0.039(5) & 4\\
    \hline
    J1814$-$1744 & 51700 & \phantom{000}5(2)  &                  &5 \\
                 & 52117 & \phantom{00}33(2) &\phantom{000}$-$0.5(4) &5 \\
                 & 53302 &\phantom{000}7(2)  &\phantom{$-$000}2(1) & 5\\
    \hline
  \end{tabular}
\end{table}

\section*{Acknowledgements}

The Westerbork Synthesis Radio Telescope is operated by ASTRON
(Netherlands Foundation for Research in Astronomy) with support from
the Netherlands Foundation for Scientific Research NWO.  We would like
to thank N.~Wang, M.~Kramer and P.~Weltevrede for helpful discussions.

\end{document}